\documentclass[fleqn,usenatbib]{mnras}

\usepackage{newtxtext,newtxmath}
\usepackage[T1]{fontenc}

\DeclareRobustCommand{\VAN}[3]{#2}
\let\VANthebibliography\thebibliography
\def\thebibliography{\DeclareRobustCommand{\VAN}[3]{##3}\VANthebibliography}

\usepackage{graphicx}	% Including figure files
\usepackage{amsmath}	% Advanced maths commands
\title[Cyclotron and Fe lines in GRO J1750-27]{Discovery of cyclotron and narrow Fe K$_{\alpha}$ lines in HMXB GRO J1750-27}

\author[P. Sharma et al.]{
Prince Sharma,$^{1}$\thanks{E-mail: princerajsharma31@gmail.com}
Chetana Jain$^{2}$\thanks{E-mail: chetanajain11@gmail.com}
and Anjan Dutta$^{1}$
\\
$^{1}$Department of Physics and Astrophysics, University of Delhi, Delhi 110007, India\\
$^{2}$Hansraj College, University of Delhi, Delhi 110007, India\\
}

\date{Accepted XXX. Received YYY; in original form ZZZ}

\pubyear{2021}

\begin{document}
\label{firstpage}
\pagerange{\pageref{firstpage}--\pageref{lastpage}}
\maketitle

\begin{abstract}
We report on timing and spectral analysis of transient Be X-ray pulsar GRO J1750-27 using the \emph{Nuclear Spectroscopic Telescope Array (NuSTAR)} observation from September 2021. This is the fourth outburst of the system since 1995. The \emph{NuSTAR} observation was performed during the rising phase of the outburst. Pulsations at a period of 4.450710(1) s were observed in the 3--60 keV energy range. The average pulse profile comprised of a broad peak with a weak secondary peak which evolved with energy. We did not find any appreciable variation in the X-ray emission during this observation. The broad-band phase-averaged spectrum is described by a blackbody, a powerlaw or Comptonization component. We report discovery of Fe K$_{\alpha}$ line at 6.4 keV along with presence of two cyclotron resonant scattering features around 36 and 42 keV. These lines indicate a magnetic field strength of $3.7_{-0.3}^{+0.1} \times 10^{12}$  and $4.4 \pm 0.10 \times 10^{12}$ G for the neutron star. We have estimated a source distance of $\sim$ 13.6--16.4 kpc based on the accretion-disc torque models.
\end{abstract}


%%%%%%%%%%%%%%%%%%%%%%%%%%%%%%%%%%%%%%%%%%%%%%%%%%

%%%%%%%%%%%%%%%%% BODY OF PAPER %%%%%%%%%%%%%%%%%%

\section{Introduction}
Be/X-ray binary (BeXRB) systems are high-mass X-ray binaries (HMXBs) which harbour a neutron star (NS) in an elliptical orbit around a rapidly rotating Be star \citep{Reig2011}. The Be star is characterized by spectral emission lines associated with the circumstellar disc formed around it \citep{Porter2003,Rivinius2013}. The capture of matter from the circumstellar disc by the NS results in transient X-ray variability. The outbursts in BeXRBs are generally classified into two classes, viz, type I (or normal) and type II (or giant) outbursts \citep{Stella1986,Okazaki2001}. 

Type I outbursts are short-lived (few days to weeks) and often occur periodically due to increased accretion of matter on to the NS during the periastron passage of the orbit. The peak X-ray luminosities during these outbursts can reach up to L$_{X} \sim 10^{37}$ erg s$^{-1}$. Type II outbursts are relatively bright with peak luminosities reaching up to $10^{38}$ erg s$^{-1}$. Type II outbursts generally last longer and do not show any correlation with the orbital phase \citep{Wilson2008}.

GRO J1750-27 (also known as AX J1749.1-2639) was discovered with the Burst and Transient Source Experiment (BATSE) on board \emph{Compton Gamma Ray Observation (CGRO)} during its 1995 outburst \citep{Wilson1995}. It is a Be X-ray transient which hosts a 4.45 s neutron star \citep{Bildsten1997}. From observations made during the peak of the outburst, \citet{Scott1997} determined a large spin-up rate of 38 pHz s$^{-1}$. They also determined an orbital period of 29.8 d, projected semi-major axis of $\sim$ 101.8 lt-s, eccentricity of about $\sim 0.36$, a periastron longitude of $\sim 206.3^\circ$, and a source distance of 18 kpc.

A second outburst was detected from GRO J1750-27 with \emph{Swift}/BAT and \emph{INTEGRAL} in 2008 \citep{Brandt2008,Krimm2008,Pottschmidt2008}. Based on the estimated intrinsic spin-up of $\dot{P} = -7.5 \times 10^{-10}$ s s$^{-1}$ during the outburst, \citet{Shaw2009} inferred a source distance between 12 and 22 kpc, and a surface magnetic field strength of $\sim 2 \times 10^{12}$ G. The 1--10 keV spectrum of GRO J1750-27 was described with an absorbed cut-off power law \citep{Pottschmidt2008}. The broad-band 5--70 keV \emph{INTEGRAL} spectrum of the source was described with a cut-off power law with photon index $\Gamma = -0.15 \pm 0.3$ and cut-off energy $E_{\rm c} = 6.0_{-0.4}^{+0.5}$ keV. Inverse Compton scattering \citep[\texttt{compTT} model,][]{Titarchuk1994} of soft thermal photons by hot plasma medium of $4.6 \pm 0.1$ keV, also explained the spectrum well \citep{Shaw2009}. However, the large calibration uncertainties and distinct instrumental responses could not provide conclusive inferences.  

\emph{Fermi}/GBM detected a giant outburst from GRO J1750-27 in 2014 \citep{Finger2014}. The follow-up observation by using \emph{INTEGRAL} reported a source flux of $7 \times 10^{-10}$ and $7.4 \times 10^{-10}$ erg cm$^{-2}$ s$^{-1}$ in 3--10 and 20--60 keV, respectively \citep{Boissay2015}. Based on the X-ray and near-infrared observations, \citet{Lutovinov2019} identified the infrared counterpart as a star of spectral type not later than B1--2 and estimated a magnetic field strength of NS to be $\approx$ (3.5--4.5) $\times 10^{12}$ G. 

\emph{Fermi}/GBM detected another outburst in GRO J1750-27 in 2021 September \citep{Malacaria2021}. In this letter, we present timing and broadband spectral study of GRO J1750-27 by analysing observations taken by \emph{Nuclear Spectroscopic Telescope Array (NuSTAR)} during this outburst. 

\section{Observations}
\emph{NuSTAR} is the first high-energy space mission by NASA, designed to achieve good sensitivity above 10 keV for imaging purposes with its advanced X-ray focusing telescopes \citep{Harrison2013}. Its payload comprise two identical detectors, known as focal plane module A (FPMA) and B (FPMB). These detectors assembled at the focal points of each of the two co-aligned, grazing incidence telescopes provide a field of view of 10 arcmin with an angular resolution of 18 arcsec at 10 keV and operates in 3--70 keV energy band. \emph{NuSTAR} has an energy resolution (full-width at half-maximum) of 400 eV and 900 eV at 10 keV and 60 keV, respectively.  

\begin{figure}
\centering
	\includegraphics[width=\columnwidth]{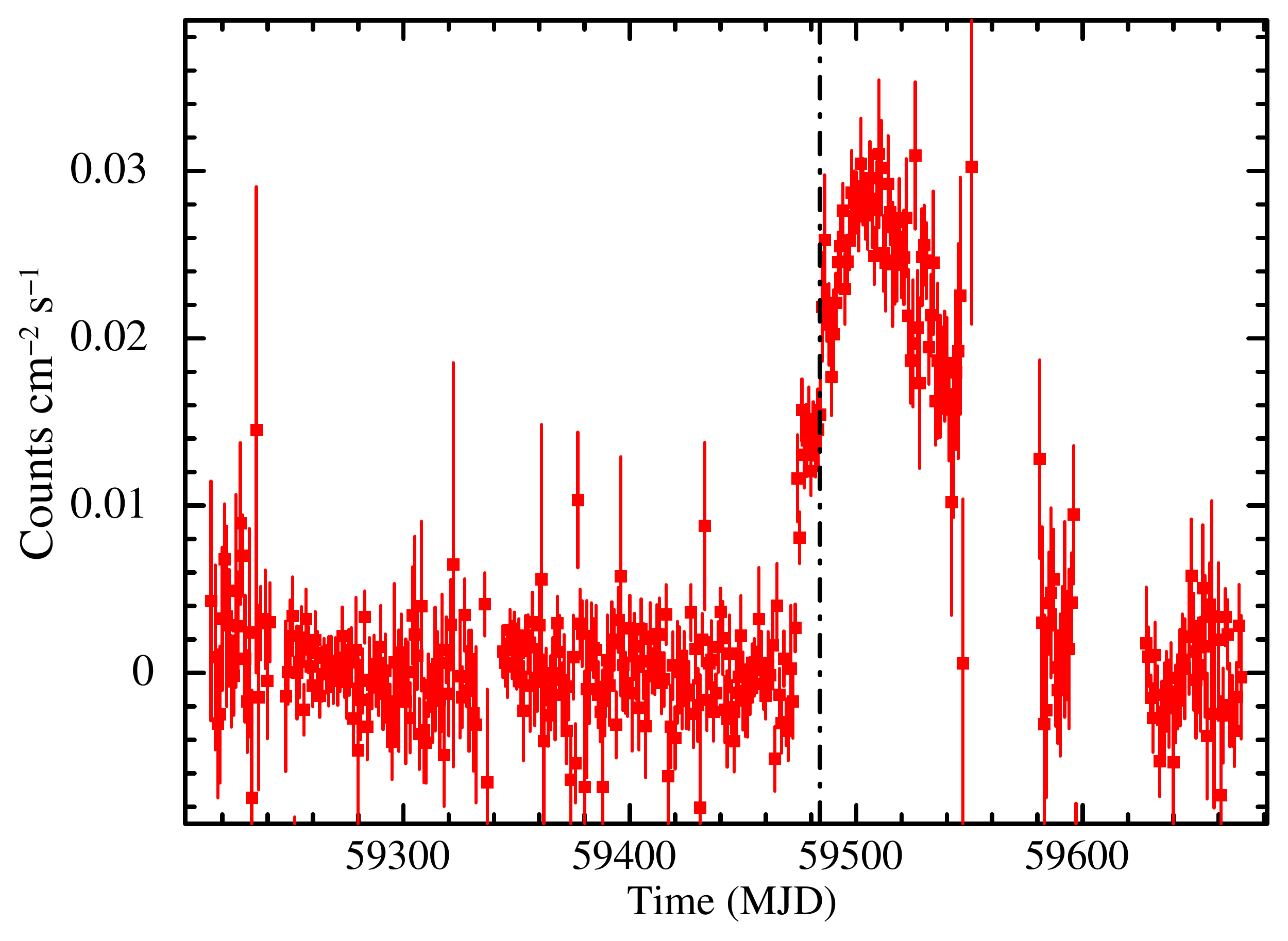}
    \caption{The daily-averaged 15--50 keV \emph{Swift}/BAT light curve of GRO J1750-27. The epoch of \emph{NuSTAR} observation is marked with the dash-dotted vertical line.}
    \label{fig:swiftlc}
\end{figure}

\emph{NuSTAR} observed GRO J1750-27 on 2021 September 27 for an exposure of 29.8 ks per module (Observation ID 90701331002). The daily-averaged long-term \emph{Swift}/BAT (15--50 keV) \citep{Krimm2013} light curve is shown in Figure~\ref{fig:swiftlc} starting from 2021 January (MJD 59215) up to 2022 March (MJD 59670). The vertical line shows the time of \emph{NuSTAR} observation (MJD 59484) used in this work. The outburst started at around MJD 59472 (2021 September 15) and \emph{NuSTAR} observation was taken about 12 d after the onset of the outburst during its increasing flux phase.

\begin{figure*}
\centering
	\includegraphics[width=\columnwidth]{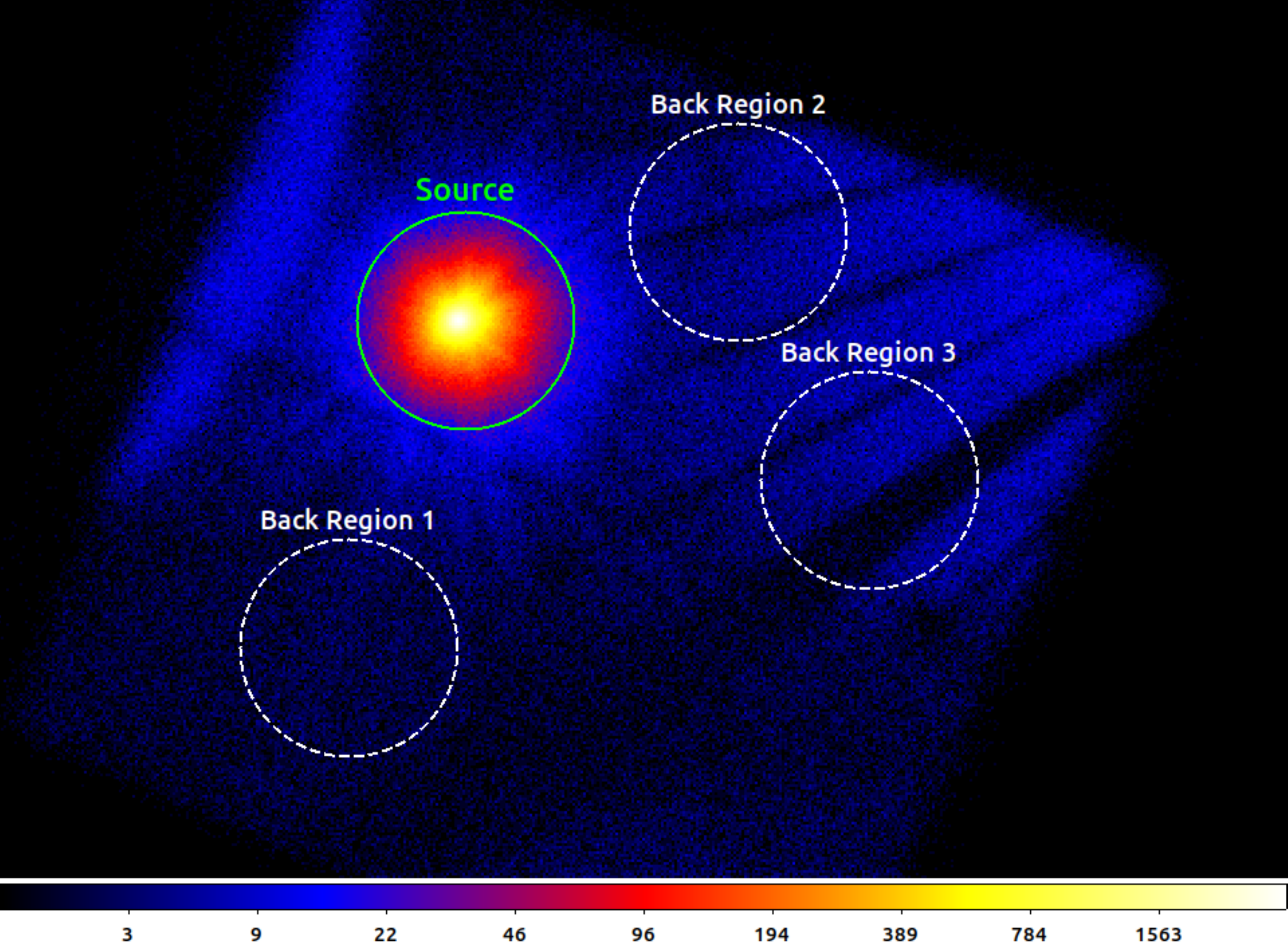}
	\includegraphics[width=\columnwidth]{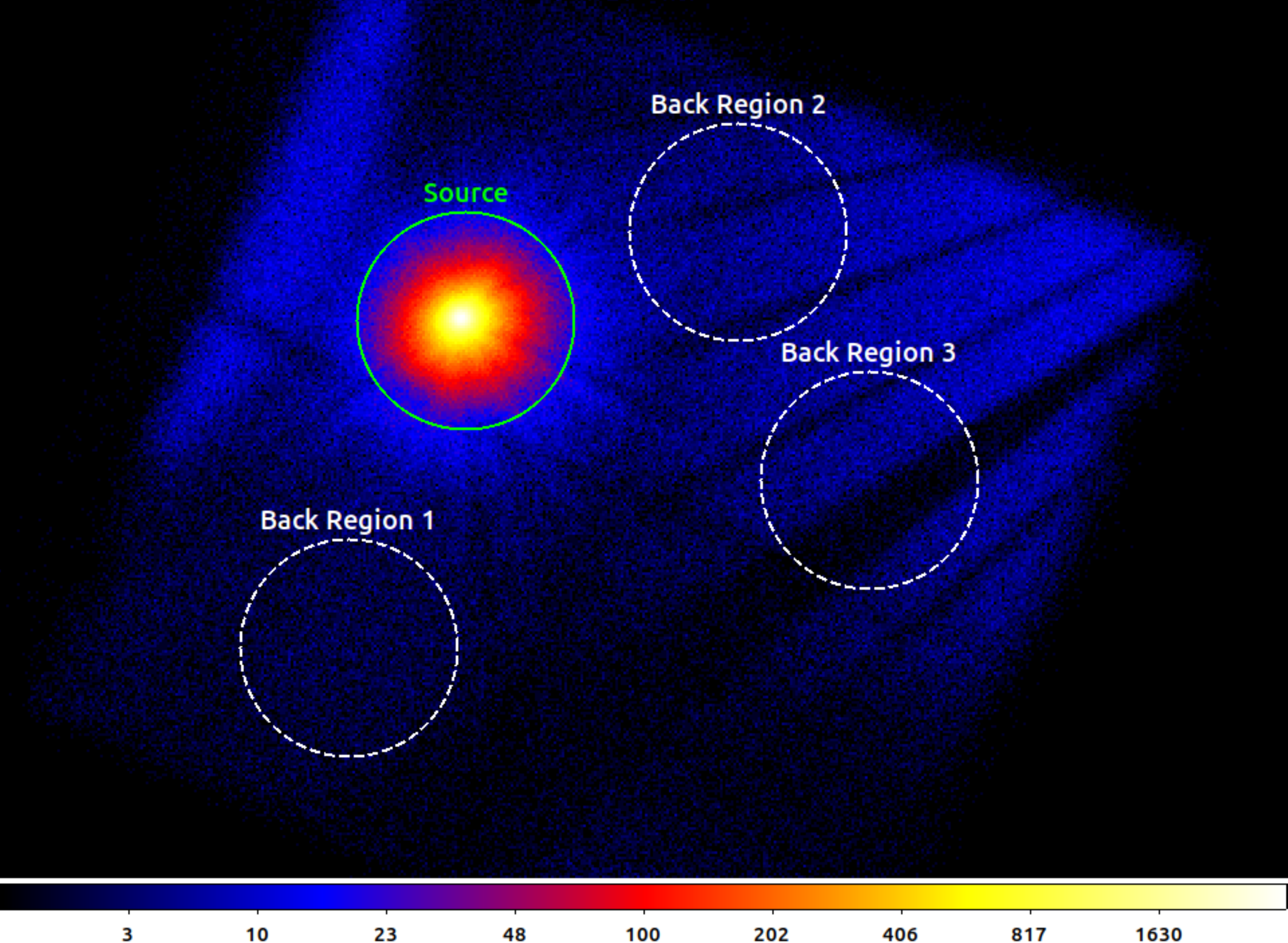}
    \caption{The images of GRO J1750-27 as observed with FPMA (left) and FPMB (right) during the 2021 September observation. The green circles mark the regions used to extract the source products. The dotted white circular region 1 represents the principal background extraction region used for the analysis while regions 2 and 3 have been used to study the effect of different background selections on the analysis. The color palette gives counts per pixel.}
    \label{fig:fov}
\end{figure*}

We used \textsc{nustardas}\_01Apr20\_v1.9.2 and \emph{NuSTAR} \textsc{caldb} version 20200912 distributed with \textsc{heasoft} v6.27 for the standard data reduction. We processed the raw event file by using the task \textsc{nupipeline} to apply standard screening and calibration and generate level 2 event files. We used a circular region with a radius of 100 arcsec centred on the source to extract the source events for both FPMA and FPMB. It is possible that during the observation, the field of view of detectors was significantly contaminated by emission from nearby sources. As shown in Figure~\ref{fig:fov}, we extracted background events from different circular regions, all having 100 arcsec radii. We applied the barycentre correction to the photon arrival time in the source event file by using the source coordinates R.A. $= 17^{\rm h} 49^{\rm m} 12.969^{\rm s}$ and Dec. $= -26^\circ 38'38.930''$ \citep{Minniti2017}. We processed the corrected level 2 event files by using the task \textsc{nuproducts} to generate the light curves, spectra, and response files for each detector. The barycenter corrected photon arrival times were corrected for orbital modulation \citep{Jenke2012} by using the ephemeris of \citet{Scott1997}.

\begin{figure}
\centering
	\includegraphics[width=\columnwidth]{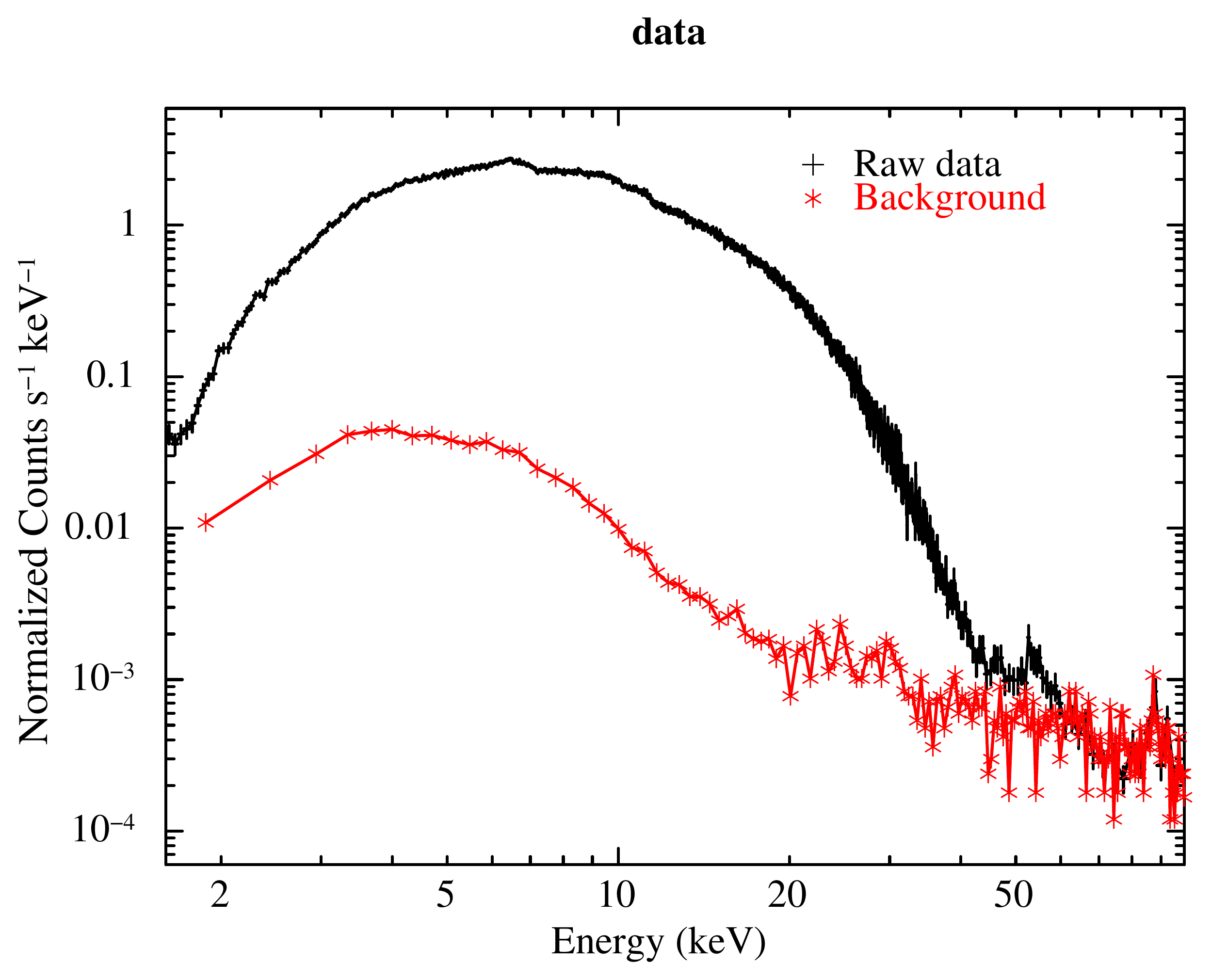}
    \caption{The source and background emission spectra of GRO J1750-27 for \emph{NuSTAR} data. The background is dominating above 60 keV.}
    \label{fig:back}
\end{figure}

The source emission was dominant up to 60 keV above background (Figure~\ref{fig:back}). We re-binned the final spectra to contain at least 25 counts per energy bin, to improve the data statistics. We have used the energy range between 3 and 60 keV for the subsequent analysis.

\section{Results}
\subsection{Timing Analysis}
\label{sec:timing} 
\begin{figure*}
\centering
	\includegraphics[width=\columnwidth]{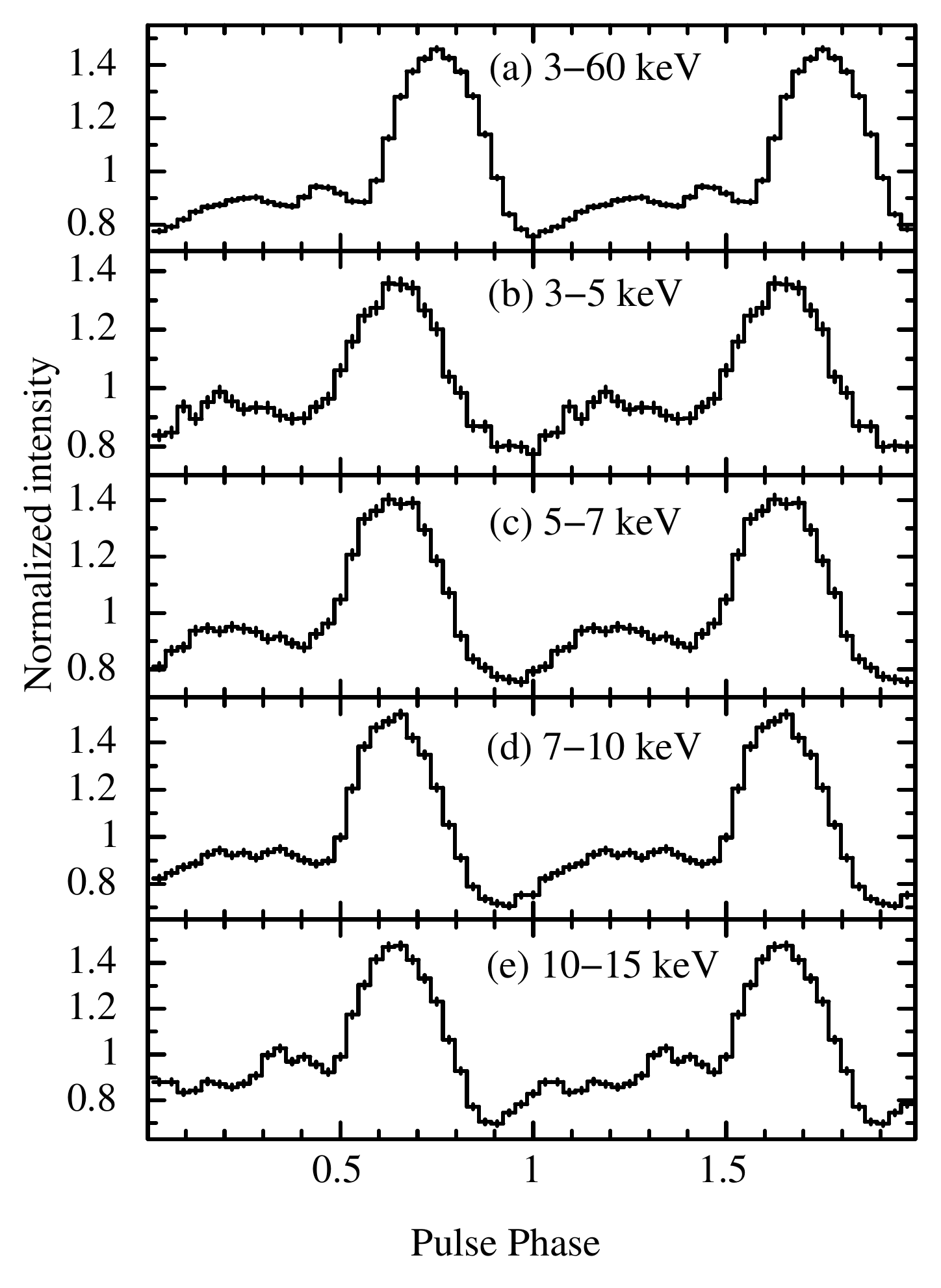}
	\includegraphics[width=\columnwidth]{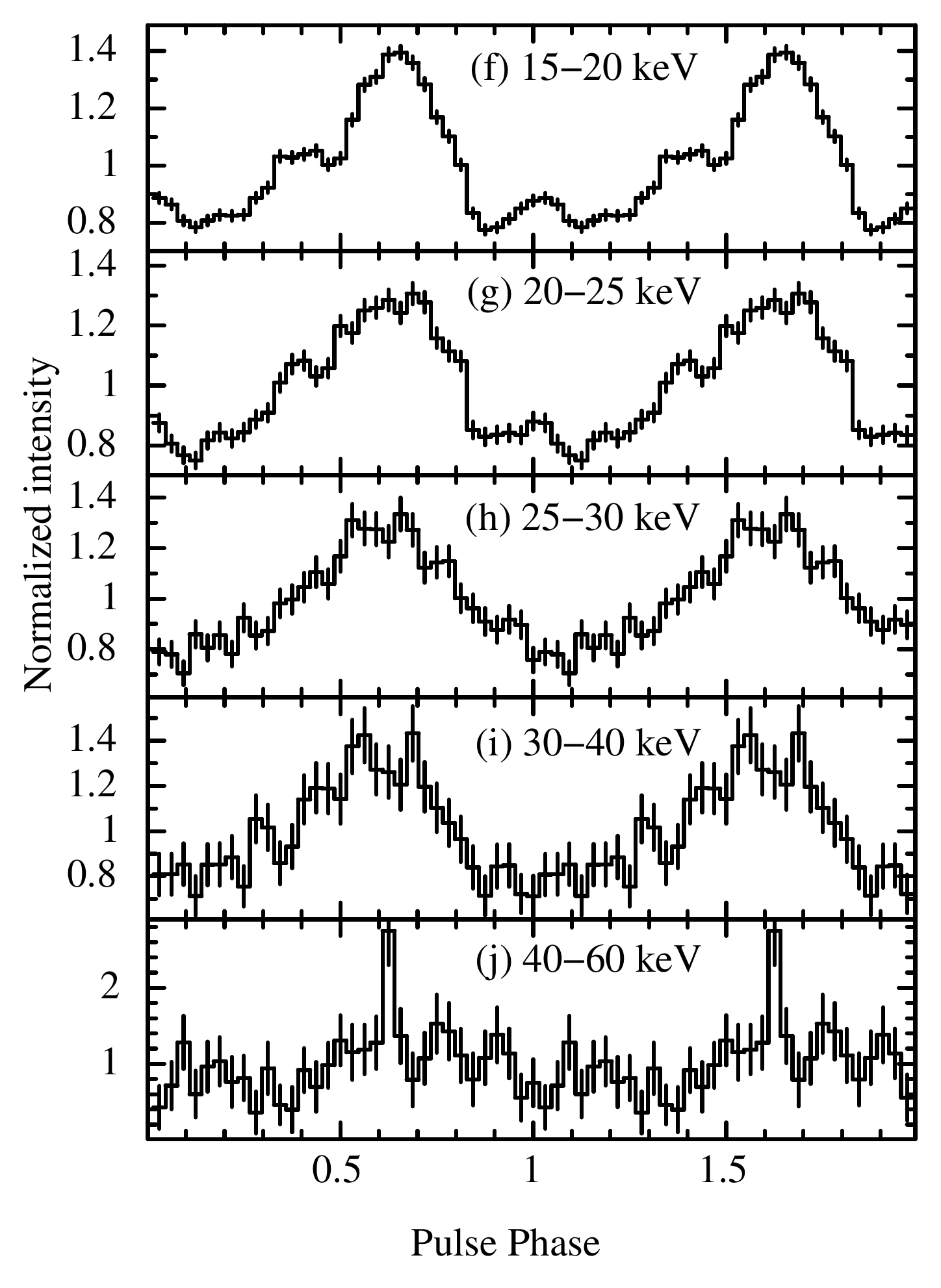}
    \caption{Energy-resolved pulse profiles of GRO J1750-27 for \emph{NuSTAR} observation folded at the best period $P = 4.450710(1)$ s. Two cycles are shown for clarity.}
    \label{fig:pulse}
\end{figure*}

We generated light curves in 3--5, 5--7, 7--10, 10--15, 15--20, 20--30, 30--40, 40--60, and 3--60 keV energy ranges with a time resolution of 0.1 s. The X-ray emission did not show any significant variation with time during the entire observation. Therefore, we have used the entire light curve for the timing analysis.

We searched the 3--60 keV light curve for coherent pulsations from the source. The power density spectrum (PDS) was generated between frequency range $6.1 \times 10^{-5}$ to 0.25 Hz by using the \texttt{ftool} \textsc{powspec}. A refined search based on chi-square maximization technique (\textsc{efsearch}) gave a best-period of 4.450710(1) s (MJD 59484.216). The uncertainty in the period determination was estimated using the bootstrap method and simulating 5000 light curves \citep{Lutovinov2012,Lutovinov2013}.
%A significant signal was detected around 0.225 Hz in the PDS.
We investigated the evolution of pulse profile with energy by using the light curves in all the above mentioned energy ranges. We folded the light curves at the best period of 4.450710(1) s. Figure~\ref{fig:pulse} shows evolution of pulse shape with energy. The pulse profile shows a significant energy dependence. They are single peaked up to 10 keV. This peak is marginally narrower in 7--10 keV compared to 3--5 and 5--7 keV profiles. A secondary peak appears above 10 keV up to 20 keV. It disappears above 20 keV and the main peak broadens above  20 keV range. The averaged 3--60 keV pulse profile shows a sharp dominating peak along with a weak secondary peak.

\subsection{Spectral Analysis}
\subsubsection{Phase-averaged spectroscopy}
\label{sec:spectral} 
We have used the spectral analysis package \textsc{xspec} v12.12.0 \citep{Arnaud1996} for the spectral fitting of \emph{NuSTAR} spectra of GRO J1750-27. Considering region 1 (Figure~\ref{fig:fov}) as the background, we performed the simultaneous spectral fitting of FPMA and FPMB spectra by introducing a cross-normalization factor. We fixed its value at 1 for FPMA and left it free to vary for FPMB. The normalization factor acquired a value of $1.01 \pm 0.01$. We have adopted the solar abundances from \citet{Wilms2000}, the photoelectric cross-sections from \citet{Verner1996} and have used the $\chi^2$ statistics as the test statistics. We have considered the 3--60 keV energy range for the phase-averaged spectral analysis. All the parameter uncertainties in the results are reported at 90 per cent confidence level.

Following the standard continuum models used to describe the HMXB spectra, we fit the phase-averaged spectrum of GRO J1750-27 with absorbed powerlaw modified with high-energy cut-off (HEPL, hereafter) and cut-off powerlaw (CPL, hereafter) models. Both the models failed to provide an acceptable fitting to the spectrum. Addition of blackbody components, \texttt{bbodyrad} (attributed to the emission from NS) or \texttt{diskbb} (to account for emission from accretion disc), improved the fitting statistically. Large positive residuals near the lower energy end were not modelled. We found an emission like feature around 6.4 keV and a broad dip between 35 and 45 keV. 

\begin{figure}
\centering
	\includegraphics[width=\columnwidth]{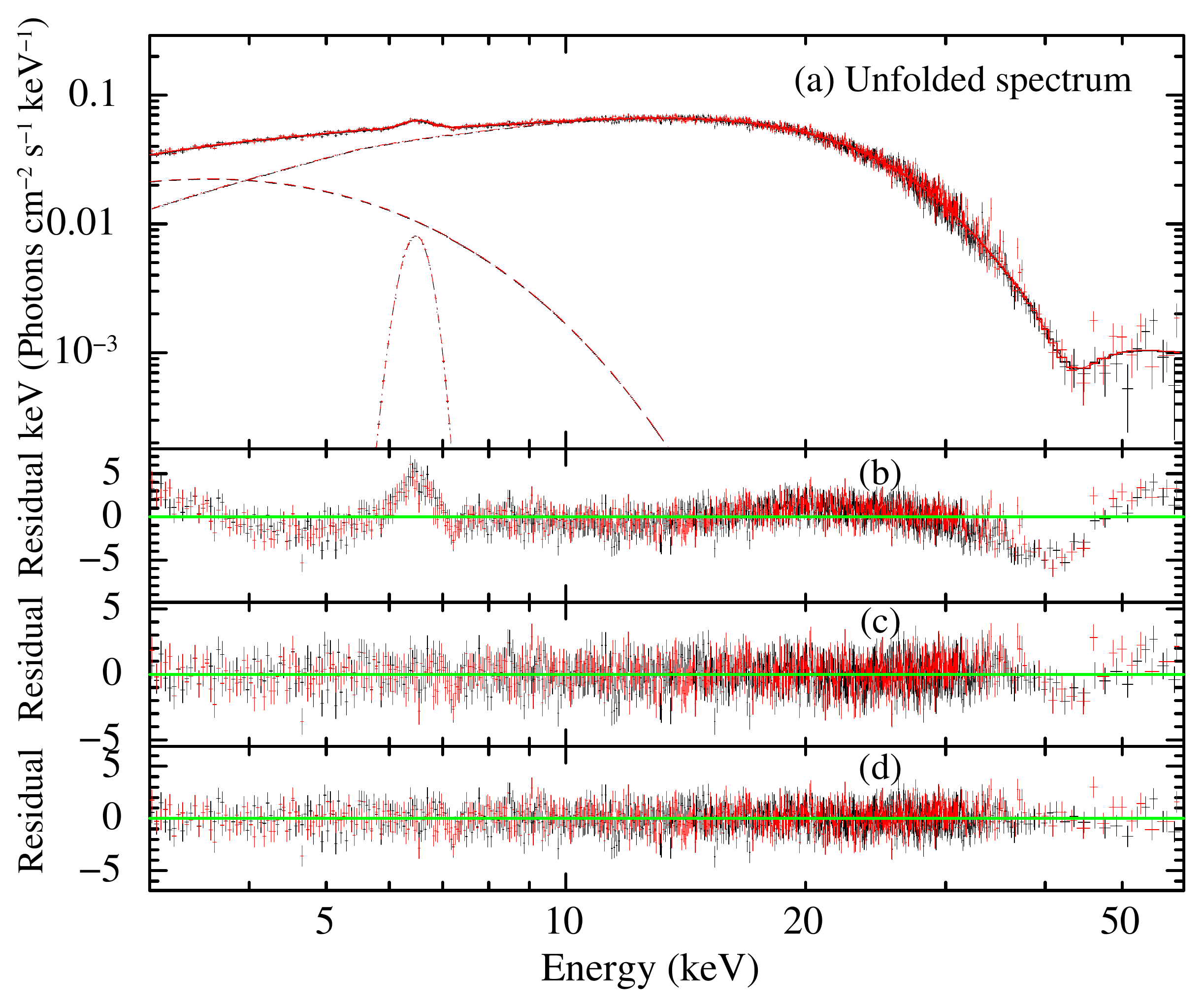}
    \caption{Best-fitting phase-averaged \emph{NuSTAR} spectra for GRO J1750-27 modelled with powerlaw model modified with high-energy cut-off. (a) Unfolded spectrum and model components. (b) Residual without \texttt{gaus} and \texttt{cyclabs} components. (c) Residual after including \texttt{gaus} and \texttt{cyclabs} components. (d) Residual after including second \texttt{cyclabs} component. The dashed, dotted, and dash-dotted lines represent \texttt{bbodyrad}, \texttt{powerlaw}, and \texttt{gaus} components, respectively.}
    \label{fig:spec1}
\end{figure}

We identified the emission feature as an iron K$_\alpha$ line and absorption feature as the Cyclotron Resonance Scattering Feature (CRSF). We used the Gaussian component \texttt{gaus} to model the emission feature and included the multiplicative component \texttt{cyclabs} to account for the absorption feature. This improved the spectral fitting significantly for both HEPL and CPL models with either \texttt{bbodyrad} or \texttt{diskbb}. Both the blackbody components provided similar fitting to the spectrum however, the best-fitting returned unreasonably higher disc temperatures ($kT_{\rm disc} \sim 2$ keV) and smaller disc radii ($R_{\rm disc} \sqrt{cos\theta} \sim$ 1--3 km, for the distance range 12--22 kpc). Thus, we dropped the models with disc blackbody components from our analysis. The models provided reasonable fitting but a narrow dip like structure persisted in the residual above 40 keV as shown in Figure~\ref{fig:spec1}(b).

We included a second \texttt{cyclabs} component to model the feature around 40 keV. The addition of second \texttt{cyclabs} to both models improved the fitting statistically with a $\Delta \chi^2$ of 37.9 and 43.6 for additional 3 degrees of freedom for the two models, respectively. The second feature was fit at $42.8 \pm 0.09$ and $42.2 \pm 0.9$ keV for HEPL and CPL, respectively. We found a broader strong feature around 36 keV along with a weaker feature around 42 keV in the spectra.

The HEPL model with \texttt{bbodyrad}, a Gaussian and two cyclotron absorption components returned a $\chi^2$ of 1623.8 for 1599 degrees of freedom. The CPL model yielded a $\chi^2$ of 1646.5 for 1599 degrees of freedom. In \textsc{xspec} the two models are defined as \texttt{tbabs*cyclabs*cyclabs*(bbodyrad + powerlaw*highecut + gaus)} (M1, hereafter) and \texttt{tbabs*cyclabs*cyclabs*(bbodyrad + cutoffpl + gaus)} (M2, hereafter). The best-fitting results obtained from the two models are summarized in Table~\ref{tab:spectral} with the corresponding best-fitting spectra with model M1 displayed in Figure~\ref{fig:spec1}.

\begin{table}
\caption{Best-fitting spectral parameters for the phase-averaged \emph{NuSTAR} spectrum of GRO J1750-27. All the errors are quoted at 90 per cent confidence level.}
\label{tab:spectral}
%\resizebox{\columnwidth}{!}{
\scalebox{0.9}{
\begin{tabular}{ c c c c} 
\hline
 \textbf{Parameters} & \textbf{Model M1} & \textbf{Model M2} & \textbf{Model M3} \\
\hline
  $C_{\rm FPMB}$ & $1.01 \pm 0.01$ & $1.01 \pm 0.01$ & $1.01 \pm 0.01$\\[0.7ex]
 
  $N_{\rm H}$ ($10^{22}$ cm$^{-2}$) &$1.7_{-0.7}^{+0.8}$ & $2.8 \pm 0.7 $ & $3.3_{-0.7}^{+0.9} $\\[0.7ex]

 $kT_{\rm BB}$ (keV) & $1.1 \pm 0.1 $  & $ 1.2 \pm 0.1$ & \\[0.5ex]
  $N_{\rm BB}$ & $12 \pm 2 $ & $ 5_{-1}^{+2}$ &  \\[0.5ex]
  ${R_{\rm BB}^a}$ (km) & $ 4.9 \pm 0.4$ & $ 3.1_{-0.3}^{+0.6}$ & \\[0.5ex]
  $f^{b}_{\rm bol}$ & $ 1.42 \pm 0.04 $ &  & \\[0.7ex]
 % & $ 1.06 \pm 0.02 $

  $\Gamma$  & $-0.6 \pm 0.4 $ & $ 0.3_{-0.5}^{+0.1}$  & $1.25_{-0.02}^{+0.01} $\\[0.5ex]
   $E_{\rm cut}$ (keV) & $ 5.5 \pm 0.2 $ & $> 0.01$ & \\[0.5ex]
   $E_{\rm fold}$ (keV) & $ 10.3_{-2.0}^{+4.0} $  & & \\[0.5ex]
   $kT_{\rm e}$ (keV) & & & $7.6_{-0.3}^{+0.9} $\\[0.5ex]
  $kT_{\rm seed}$ (keV) & & & $< 0.5$ \\[0.5ex]
  Norm ($10^{-3}$) & $2.7_{-1.1}^{+2.7} $  & $ 15.5_{-0.1}^{+2.9}$ & $6.1_{-2.1}^{+0.3} $ \\[0.7ex]
  $f^{b}_{\rm bol}$ & $ 38.90 \pm 0.01 $ &  & $42.8 \pm 0.2 $ \\[0.7ex]

  $E_{\rm line}$ (keV) & $6.47 \pm 0.03 $ & $6.43 \pm 0.03 $ & $6.44 \pm 0.03$ \\[0.5ex]
 $\sigma$ (keV) & $ 0.25_{-0.03}^{+0.04} $ & $0.28 \pm 0.04 $ & $0.26 \pm 0.04 $ \\[0.5ex]
  Norm ($10^{-4}$) & $8.22_{-1.01}^{+1.04} $ &$9.2 \pm 0.9$ & $7.87_{-0.80}^{+0.86} $ \\[0.5ex]
   EW (eV) &$93_{-14}^{+55}$ & $< 1000 $ & $84_{-10}^{+8} $ \\[0.7ex]
  $f^{c}_{\rm bol}$ & $ 8.5 \pm 0.6 $ &  & $8.1 \pm 0.5 $\\[0.7ex]

 $E_{\rm cyc1}$ (keV) & $36.0_{-2.5}^{+0.9} $ & $ 35.4_{-1.5}^{+0.8}$ & $ 33.9_{-2.1}^{+1.3}$ \\[0.5ex]
 Width (keV) & $17.7_{-7.3}^{+2.9} $ & $ 25.6_{-2.5}^{+2.3}$ &  $ 18.5_{-4.6}^{+3.0}$\\[0.5ex]
 Depth & $ 2.1_{-1.4}^{+0.5}$ & $ 3.4 \pm 0.2$ & $ 1.8_{-0.9}^{-0.3}$ \\[0.7ex]

 $E_{\rm cyc2}$ (keV) & $42.8 \pm 0.9 $ & $42.2 \pm 0.9$ & $ 42.5 \pm 0.8$ \\[0.5ex]
  Width (keV) & $4.2_{-1.9}^{+6.6} $ & $ 4.4_{-1.7}^{+2.1}$ & $ 5.6_{-2.2}^{+4.1}$\\[0.5ex]
 Depth & $ 0.9_{-0.3}^{+0.9}$ & $ 0.8 \pm 0.2 $ & $ 1.2_{-0.3}^{+1.0}$ \\[0.7ex]

  & & & \\
 $f^{b}_{\rm Total}$ &$ 40.40 \pm 0.02 $ &  & $42.9 \pm 0.2  $ \\[0.7ex] 

\hline
 $\chi^2$/$\nu$ & 1623.8/1598 & 1646.5/1599 & 1634.9/1600 \\
\hline

 \multicolumn{4}{l}{Model M1 = \texttt{tbabs*cyclabs*cyclabs*(bbodyrad +po*highecut + gaus)}} \\
  \multicolumn{4}{l}{Model M2 = \texttt{tbabs*cyclabs*cyclabs*(bbodyrad + cutoffpl + gaus)}} \\
  \multicolumn{4}{l}{Model M3 = \texttt{tbabs*cyclabs*cyclabs*(nthcomp[bb] + gaus)}} \\
  \multicolumn{4}{l}{$^{a}$ Parameters are calculated for a source distance of 14 kpc.}\\
  \multicolumn{4}{l}{$^{b} f$ is the unabsorbed bolometric flux in 3.0--60.0 keV in $10^{-10}$ erg cm$^{-2}$ s$^{-1}$.}\\ 
  \multicolumn{4}{l}{$^{c} f$ is the unabsorbed bolometric flux in 3.0--60.0 keV in $10^{-12}$ erg cm$^{-2}$ s$^{-1}$.}
\end{tabular}}
\end{table}

\begin{figure}
\centering
	\includegraphics[width=\columnwidth]{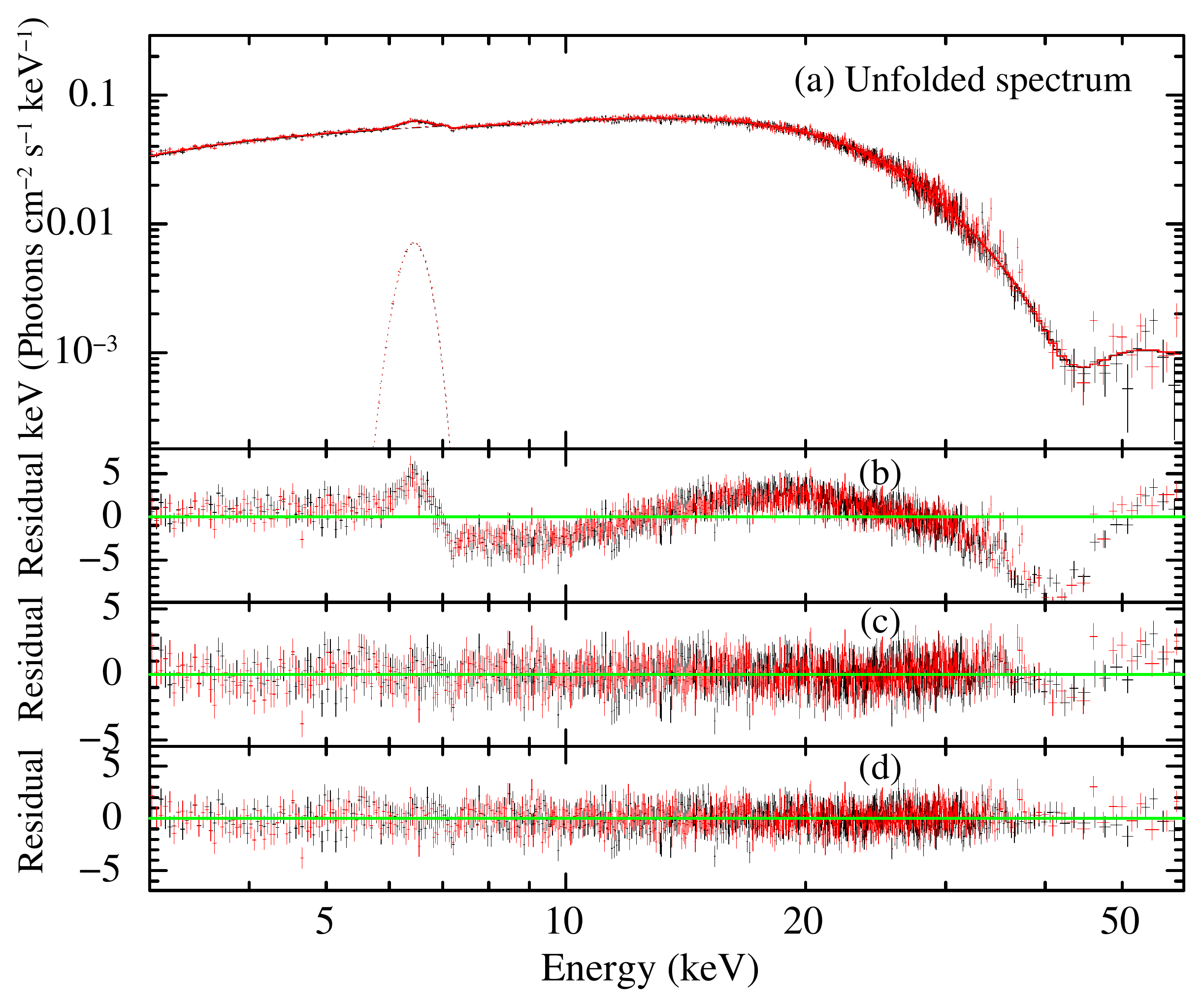}
    \caption{(a) Unfolded \emph{NuSTAR} spectrum and model components modelled with \texttt{tbabs*cyclabs*cyclabs*(nthcomp[bb] + gaus)}. (b) Residual without \texttt{gaus} and \texttt{cyclabs} components. (c) Residual after including \texttt{gaus} and \texttt{cyclabs} components. (d) Residual after including second \texttt{cyclabs} component. The dashed and dash-dotted lines represent \texttt{nthcomp} and \texttt{gaus} components, respectively.}
    \label{fig:spec2}
\end{figure}

For the completeness of this work, we fit the spectra of GRO J1750-27 by using the thermal Comptonization model \texttt{nthcomp} \citep{Zdziarski1996,Zycki1999} as suggested by \citet{Shaw2009}. Similar to models M1 and M2, a Gaussian and two cyclotron absorption components were also required to model the emission and cyclotron absorption features. The addition of \texttt{bbodyrad} or \texttt{diskbb} to account for the thermal emission at lower energy with either seed photon source did not improve the fitting significantly thus, we removed these components. The model \texttt{tbabs*cyclabs*cyclabs*( nthcomp[bb] + gaus)} (M3, hereafter) provided an acceptable fitting with a $\chi^2$ of 1634.9 for 1600 degrees of freedom. Figure~\ref{fig:spec2} shows the best-fitting spectrum with model M3 and the resulting spectral parameters reported in Table~\ref{tab:spectral}.

We also tried the Gaussian absorption component \texttt{gabs} to model the CRSF features. The fitting with \texttt{gabs} provided identical results in terms of spectral parameters and $\chi^2$. The best fits returned the two CRSF energies as $43.9_{-1.9}^{+0.6}$, $42.7_{-1.0}^{+1.2}$ keV; $51.2 \pm 0.05$, $42.8_{-1.3}^{+1.9}$keV ; and $44.8 \pm 0.5$, $42.5 \pm 1.1 $ keV with M1, M2, and M3, respectively. Clearly, the centroid energy from the \texttt{gabs} model shows a systematic shift as compared to that obtained from the \texttt{cyclabs} model. Similar shifts have been reported in several other sources \citep[see e.g.,][]{Staubert2014,Furst2015,Tsygankov2016,Doroshenko2017}. While the goodness of fit is similar for both the components, our results show that the CRSF parameters are consistent across three models with \texttt{cyclabs} component. Therefore, we have dropped the \texttt{gabs} component from our analysis.
% weak feature around 42 keV were in agreement with results obtained with \texttt{cyclabs} while best-fitting energies In general, the energy shifts up to $\sim $ 4 keV have been observed but the difference is slightly larger in our case \citep{Lutovinov2015,Tsygankov2016}.

\begin{figure}
\centering
	\includegraphics[width=\columnwidth]{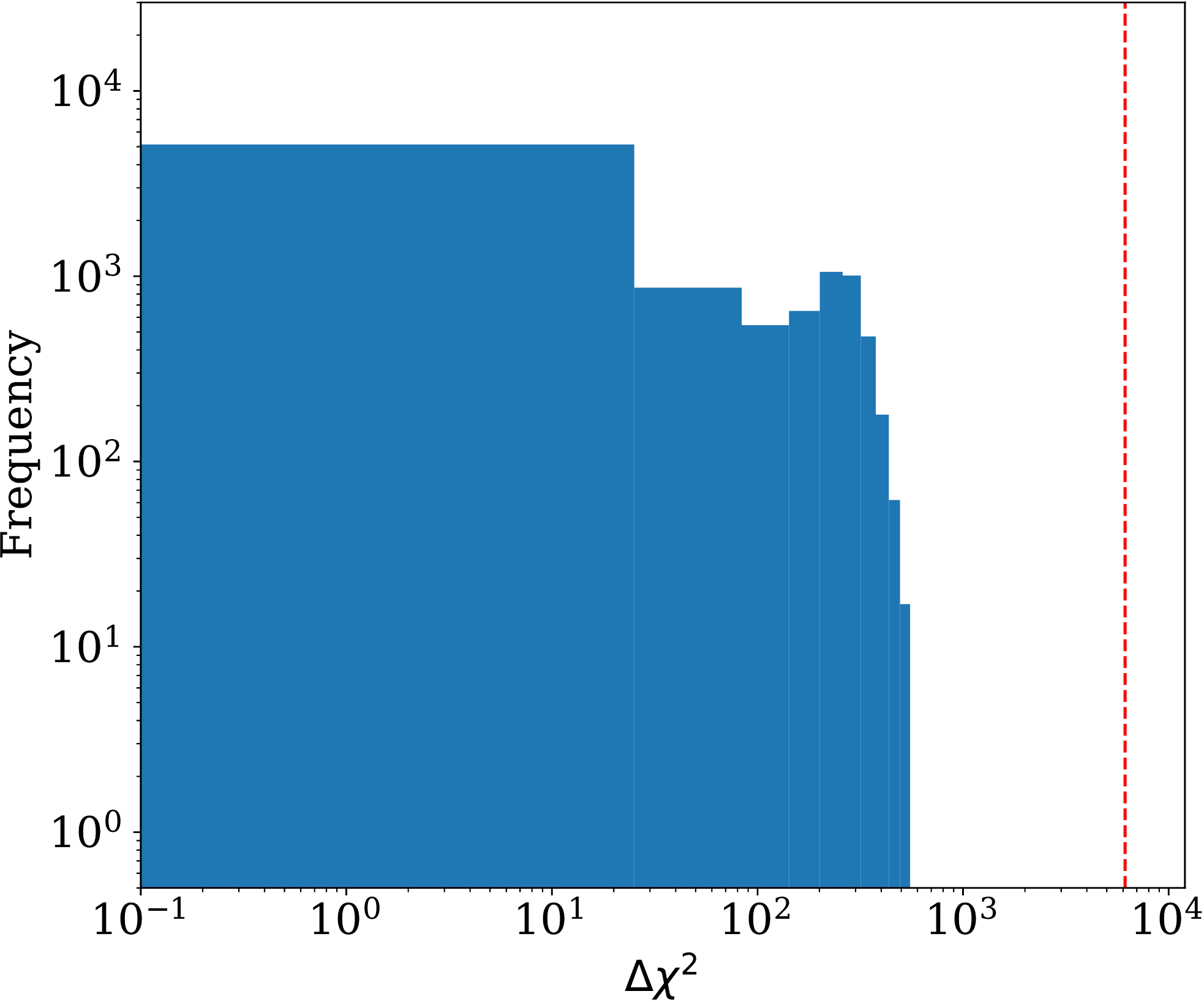}
	\includegraphics[width=\columnwidth]{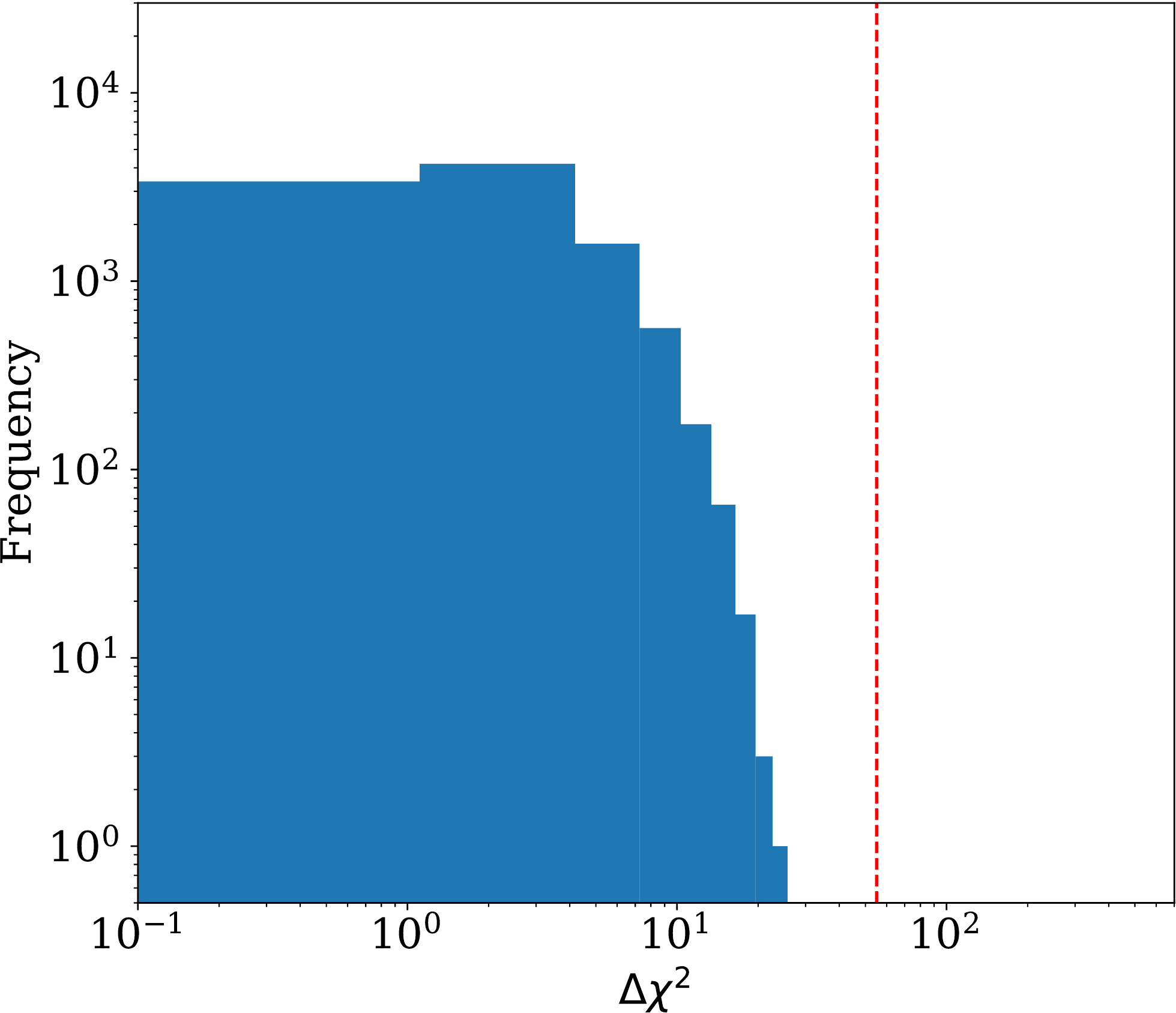}
    \caption{The histograms showing the results of the $10^4$ \texttt{simftest} simulations to test the significance of \texttt{cyclabs} features with model M3. Top and bottom plots correspond to the simulations for the first and second cyclotron features, respectively. The vertical dashed lines mark the observed $\Delta \chi^2$ for two cases.}
    \label{fig:simftest}
\end{figure}

We have found the absorption features in the residuals with all the three models. The inclusion of cyclabs component improved the fitting statistically with most significant improvement for model M3 with $\Delta \chi^2$ of 6138.4 and 55 for the first and second features, respectively. We have utilized the \texttt{simftest} routine from \textsc{xspec} to test the significance of the cyclotron lines in the spectra with model M3. It uses Monte Carlo simulations to generate simulated data-sets based on the original data and use these to estimate $\Delta \chi^2$ for additional components. We ran $10^4$ simulations for each of the two additional \texttt{cyclabs} components. The results are plotted in Figure~\ref{fig:simftest}. The large difference of $\Delta \chi^2$ between the observed and maximum value from simulated data-sets for the two features confirms the significant detection of these features in the spectra.

It is worthwhile to mention here the effectiveness of the choice of background region on the estimation of spectral parameters. We performed the simultaneous spectral fitting of FPMA and FPMB spectra for regions 2 and 3 (Figure~\ref{fig:fov}). The best-fitting spectral parameters were consistent within errors for all three cases. About 10 per cent energy shift in centroid line energy of 36 keV feature was seen for model M1 (region 1 to region 3). It was about 2 per cent for M2 and 6 per cent for M3. The second feature at 42 keV did not show significant variation.

We used the convolution model \texttt{cflux} to estimate the unabsorbed flux in the energy range 3.0-60.0 keV for models M1 and M3 and their components. The flux measurements for model M2 were not constrained due to poorly constrained $E_{\rm cut}$ thus, we do not report flux for M2.

\subsubsection{Phase-resolved spectroscopy}

\begin{figure}
\centering
	\includegraphics[width=\columnwidth]{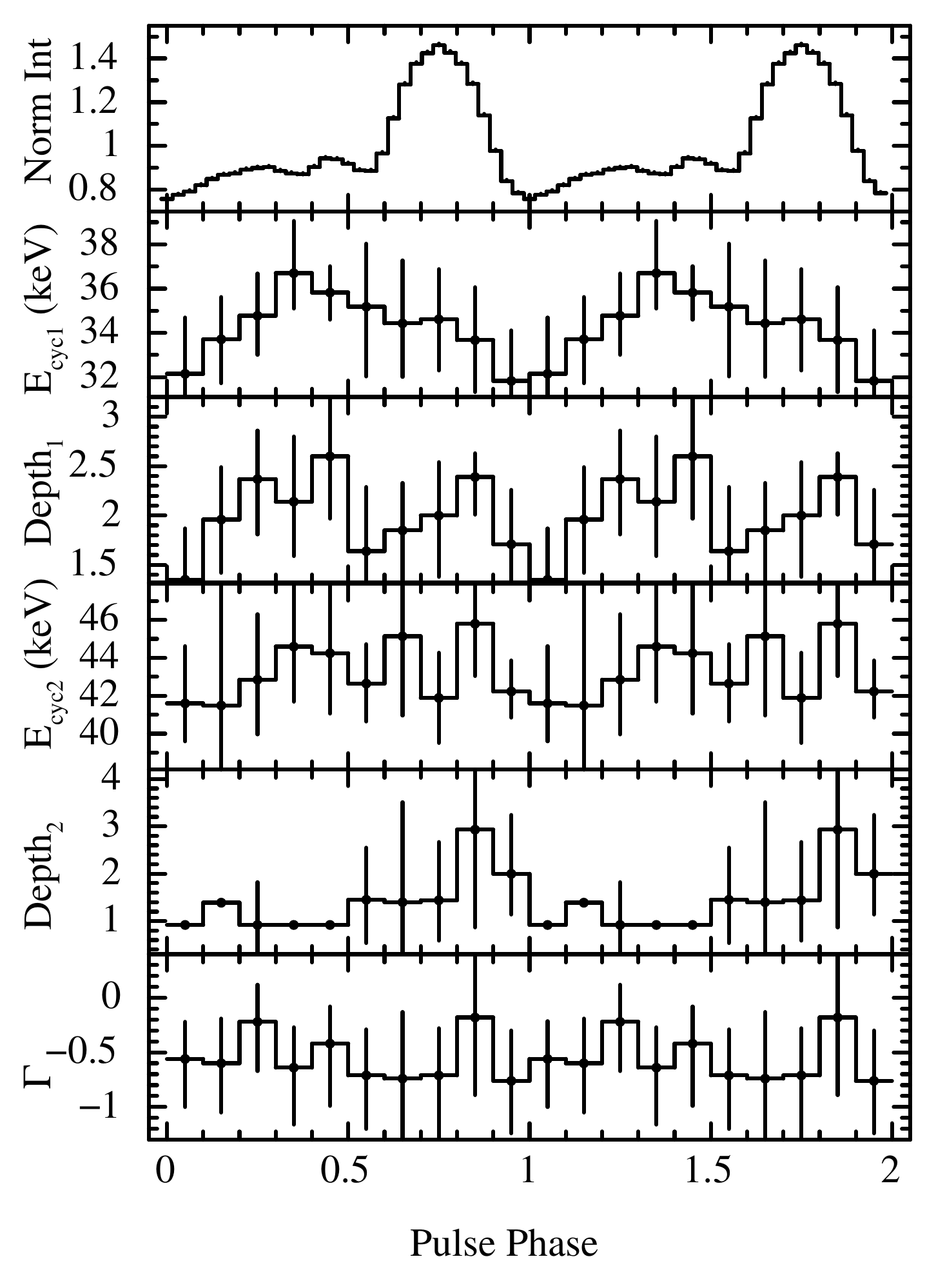}
    \caption{Variation of the spectral parameters of GRO J1750-27 obtained with the phase-resolved spectroscopy by using the \emph{NuSTAR} observation. The top panel presents the FPMA pulse profile in 3--60 keV. The errors are plotted for 90 per cent confidence level.}
    \label{fig:phase_spec}
\end{figure}
We performed the phase-resolved spectroscopy in an attempt to study the evolution of CRSF parameters with the pulse phase of the pulsar. We generated the phase-resolved spectra in ten equal pulse-phase bins. We performed the analysis in 3--60 keV energy range and used the corresponding background and response files. We performed the simultaneous spectral fitting of FPMA and FPMB spectra with all three models. While all the models returned consistent best-fitting parameter values, here we report the results with model M1 only. Due to the limited statistics after resolving the spectra, we fixed the $N_{\rm H}$, Fe K$_{\alpha}$ line parameters and width of the cyclotron lines at the phase-averaged values for each phase bin during the spectral fitting. We found the absorption like features between 35--45 keV in the residuals for all the phase bins. We used the \texttt{cyclabs} components to model the features. However, the depth for second feature was not constrained for phase bins 0--0.1, 0.3--0.4, and 0.4--0.5, so fixed the value at the phase-averaged value for fitting.

Figure~\ref{fig:phase_spec} shows the variation of the cyclotron line parameters and photon index with the pulse phase. For the strong feature (CRSF$_1$), the line energy varies significantly from around 32 keV near the pulse minimum to around 37 keV close to the weaker secondary pulse peak. It decreases gradually after that. The depth of line varies differently and peaks near the dips in the pulse profiles. The line energy for CRSF$_2$ does not show significant variation and is roughly constant while because of the low statistics, the variation in the line depth is not discernible. The photon index evolves marginally over phase.

\section{Discussion}
We present the timing and broad-band spectral properties of GRO J1750-27 from \emph{NuSTAR} observations performed during the 2021 outburst. We discovered an emission line along with two close cyclotron absorption lines in the \emph{NuSTAR} spectrum of GRO J1750-27.   

\begin{figure}
\centering
	\includegraphics[width=\columnwidth]{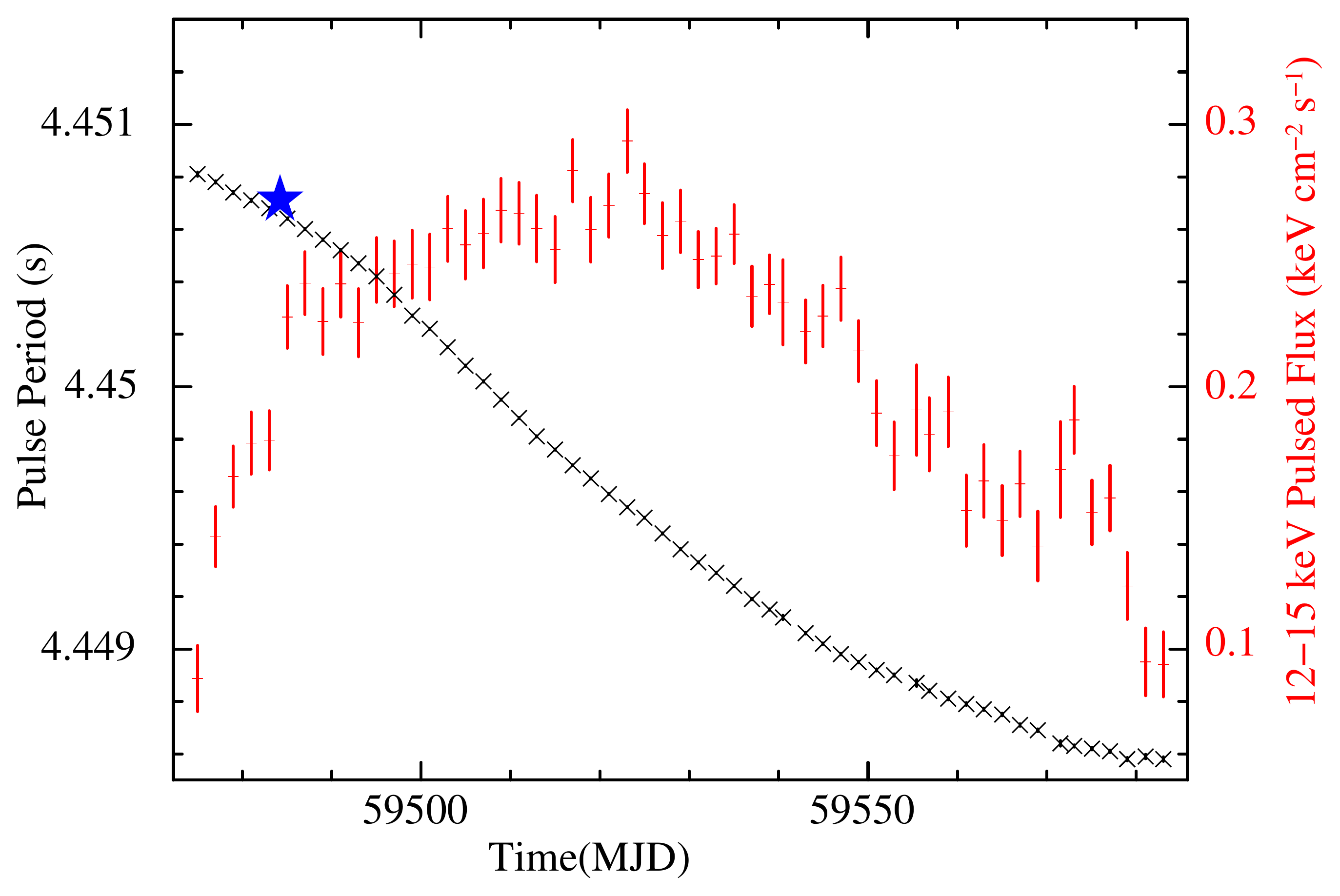}
    \caption{\emph{Fermi}/GBM monitoring of evolution of pulse period and pulsed flux of GRO J1750-27 during the outburst. The blue $*$ corresponds to the pulse period estimated in the current work.}
    \label{fig:spin}
\end{figure}

We studied the timing properties of the source. We detected pulsations at 4.450710(1) s up to 60 keV. \emph{Fermi}/GBM monitoring\footnote{https://gammaray.nsstc.nasa.gov/gbm/science/pulsars/lightcurves/groj1750.html} reveals a continuous spin-up of the source since the beginning of the outburst (Figure~\ref{fig:spin}). We inferred a spin-up rate of $-4.4 \times 10^{-10}$ s s$^{-1}$ between MJD 59472 and 59570 by using the \emph{Fermi}/GBM data. A spin-up of the similar order was reported during the 2008 outburst by \citet{Shaw2009}.

We studied the energy resolved pulse profiles of GRO J1750-27. The average profile is double peaked with a dominant primary and a weaker secondary peak, unlike the single broad peak reported for 20-70 keV BATSE profile during 2008 outburst \citep{Shaw2009}. \citet{Brandt2008} and \citet{Pottschmidt2008} reported double peaked profiles with a dominant asymmetric peak and a relatively symmetric secondary peak. We computed the pulsed fraction defined as $(I_{max} - I_{min}) / (I_{max} + I_{min})$ for energy resolved profiles. The pulsed fraction initially increases from 27 per cent for 3--5 keV to 37 per cent for 7--10 keV. Above 10 keV, the pulsed fraction decreases to 29 and 27 per cent for 15--20 and 20--25 keV energy bands, respectively. The pulsed fraction again increases to 34 per cent for 30--40 keV range above which the poor statistics limit the estimation of pulsed fraction.

From the broad-band phase-averaged spectral analysis of GRO J1750-27, we report the discovery of an Fe K$_{\alpha}$ line and presence of two CRSFs around 36 and 42 keV. Both the features were detected with considerable significance in the spectra. The X-ray continuum in 3-60 keV can be described with either an absorbed powerlaw with high-energy cut-off and a blackbody component or a thermal Comptonization model along with a 6.4 keV Fe emission line and two CRSF components. The best-fitting returned NS blackbody temperatures of $1.1 \pm 0.1$ and $1.2 \pm 0.1$ keV with inferred emission radii of $12 \pm 2$ and $5_{-1}^{+2}$ km for M1 and M2, respectively. The inferred small radii imply that emission originates from a smaller region of the NS surface/boundary layer \citep{Escorial2019}. The discovery of narrow neutral iron emission line ($\sigma < 0.3$ keV) implies that the emission line originates in a sparse and cold region, away from the NS.      

The detection of multiple CRSFs have been reported in the X-ray spectra of several accreting HMXBs \citep[e.g., 4U 0115+634, Vela X-1, Cep X-4;][]{Ferrigno2011, Maitra2013,Furst2015}. In general, these features are harmonically spaced with the ratio between the first harmonic and fundamental line of the order $\sim$ 2. However, for GRO J1750-27 we found a ratio of the order 1, too small to associate the second feature as a harmonic. We do not find any hints for the presence of any feature at lower energies ruling out the possibility that two features are rather higher harmonics. A similar detection of two close CRSFs have been reported for HMXB GX 301-2 at 35 and 51 keV \citep{Furst2018}. It is possible that the two CRSFs originate at the different altitudes from the same accretion column. Based on the theoretical predictions, the possibility of presence of only single absorption line with distorted shape (Gaussian for their case) can also explain the need of second component to model the complex shape of the feature \citep{Schonherr2007,Schwarm2017}. It is possible that two CRSFs in GRO J1750-27 spectra are formed at two different heights and are fundamental lines with corresponding magnetic field strengths at the line forming regions.

We present the first detailed phase-resolved spectroscopy of GRO J1750-27 during the outburst. The CRSF parameters show significant variation with pulse phase. The changes in the line energies with pulse phase are of the order 15 and 10 per cent for the 36 and 42 keV features, respectively. Based on the several numerical simulations for the cyclotron line parameters, a change in the line parameter values within 10--20 per cent range with pulse phase can be attributed to the changes in the viewing angle or local distortion in the NS magnetic field geometry \citep{Schonherr2007,Mukherjee2012}. The 42 keV feature varies marginally over phase implying a weaker or no dependence on the viewing angle and pulse phase.

CRSFs are a prominent part of spectra of HMXBs with magnetic field of the order $\sim 10^{12}$ G. Detection of such features allow direct estimate of the magnetic field strength of the NS with the centroid energy given by, 
\begin{equation}
 E_{\rm CRSF} = 11.6 \ \rm keV \times (1+z)^{-1} \times {B_{12}} 
\label{eq:tau}
\end{equation}
where $B_{12}$ is the magnetic field strength in units of $10^{12}$ G and $z$ is the gravitational red-shift. We have found two CRSFs in the phase-averaged spectra of GRO J1750-27 at $36.0_{-2.5}^{+0.9}$ and $42.8 \pm 0.9$ keV. The features are present irrespective of the model choice. Using the best-fitting CRSF line energies, we estimate the magnetic field strengths of $3.7_{-0.3}^{+0.1}$ and $4.4 \pm 0.1 \times 10^{12}$ G for the two features, respectively with $z \sim 0.2$ for a typical NS \citep{Lindblom1983}. 

We have utilized the \emph{Fermi}/GBM data during the outburst to estimate the spin-up of the pulsar. Based on the standard accretion-disc torque model by \citet{Ghosh1979}, the rapid spin-up during an outburst is proportional to the rate of angular momentum transfer which is proportional to the X-ray luminosity of the source. Using the estimated magnetic field strengths, $\dot{P} = 13.8$ ms yr$^{-1}$, and the measured bolometric flux in 3--60 keV for M1 (Table~\ref{tab:spectral}), equation (15) from \citet{Ghosh1979} gives source distance of $13.6 \pm 0.1$ and $14.0 \pm 0.1$ kpc. There are various other versions of this model where different values of the dimensionless angular momentum $n(\omega_s)$ are considered to describe interaction of accretion disc with the magnetosphere. For example, following the model of \citet{Wang1995} we estimated a source distance of $16.4 \pm 0.2$ and $15.9 \pm 0.1$ kpc. And employing the approach of \citet{Kluzniak2007} we have estimated a source distance of $13.9 \pm 0.2$ and $13.6 \pm 0.1$ kpc. All these numbers are consistent with the previous estimate of source distance \citep{Shaw2009}.

%According to this model the spin-up rate is given by equation
%\begin{equation}
 %-\dot{P} = 5 \times 10^{-5} \  \mathrm{s} \ \mathrm{yr^{-1}} \ M_{1.4}^{-3/7} R_{6}^{6/7} I_{45}^{-1} \times n(\omega_{s}) \ \mu_{30}^{2/7} \ (P\ L_{37}^{3/7})^{2}, 
%\label{eq:distance}
%end{equation}
%where $\dot{P}$ is in units of s yr$^{-1}$, $M_{1.4}$ is NS mass in units of 1.4 M$_{\sun}$, $R_6$ is the radius of NS in units of $10^6$ cm, $I_{45}$ is the moment of inertia in units of $10^{45}$ g cm$^{2}$, $n(\omega_{s})$ is the dimensionless torque which is taken to be $\approx 1.4$, $\mu_{30}$ is the magnetic moment in $10^{30}$ G cm$^{3}$, $P$ is the spin period in s, and $L_{37}$ is the bolometric X-ray luminosity of the source in units of $10^{37}$ erg s$^{-1}$.

\section*{Acknowledgements}

We have used the \emph{NuSTAR} data archived by the HEASARC online service maintained by the Goddard Space Flight Center. This work has made use of the \emph{NuSTAR} Data Analysis Software (NuSTARDAS) jointly developed by the ASI Space Science Data Center (SSDC, Italy) and the California Institute of Technology (Caltech, USA). AD acknowledges the support received from the IoE, University of Delhi as FRP grant. PS acknowledges the financial support from the Council of Scientific \& Industrial Research (CSIR) under the Senior Research Fellowship (SRF) scheme. 
%%%%%%%%%%%%%%%%%%%%%%%%%%%%%%%%%%%%%%%%%%%%%%%%%%
\section*{Data Availability}
The \emph{NuSTAR} data used for this work are available at the High Energy Astrophysics Science Archive Research Center (HEASARC) online service and can be accessed at \url{https://heasarc.gsfc.nasa.gov/cgi-bin/W3Browse/w3browse.pl}.
The \emph{Fermi}/GBM period history during the outburst can be accessed at \url{https://gammaray.nsstc.nasa.gov/gbm/science/pulsars/lightcurves/groj1750.html}.
The \emph{Swift}/BAT daily averagd light curve for GRO J1750-27 is available at \url{https://swift.gsfc.nasa.gov/results/transients/weak/AXJ1749.1-2639/}.



\bibliographystyle{mnras}
\bibliography{GRO1750} % if your bibtex file is called example.bib

%\begin{thebibliography}{99}
%\end{thebibliography}

%%%%%%%%%%%%%%%%%%%%%%%%%%%%%%%%%%%%%%%%%%%%%%%%%%

%%%%%%%%%%%%%%%%% APPENDICES %%%%%%%%%%%%%%%%%%%%%

%\appendix


%If you want to present additional material which would interrupt the flow of the main paper,it can be placed in an Appendix which appears after the list of references.

%%%%%%%%%%%%%%%%%%%%%%%%%%%%%%%%%%%%%%%%%%%%%%%%%%

% Don't change these lines
\bsp	% typesetting comment
\label{lastpage}
\end{document}